\documentclass[aps,prl,longbibliography,superscriptaddress,reprint]{revtex4-2}

\usepackage{graphicx}
\usepackage{amsmath}
\usepackage{multirow} 
\usepackage{physics} 
\usepackage{braket}
\usepackage{subfigure}   
\usepackage{floatrow}
\usepackage{verbatim}
\usepackage{color}
\usepackage[colorlinks,urlcolor=blue,citecolor=blue,linkcolor=blue]{hyperref}

\begin{document}

\title{Observation of spin-tensor induced topological phase transitions of triply degenerate points with a trapped ion}

\author{Mengxiang Zhang}\email{These authors contribute equally.}
\affiliation{CAS Key Laboratory of Microscale Magnetic Resonance and School of Physical Sciences, University of Science and Technology of China, Hefei 230026, China}
\affiliation{CAS Center for Excellence in Quantum Information and Quantum Physics, University of Science and Technology of China, Hefei 230026, China}

\author{Xinxing Yuan}\email{These authors contribute equally.}
\affiliation{CAS Key Laboratory of Microscale Magnetic Resonance and School of Physical Sciences, University of Science and Technology of China, Hefei 230026, China}
\affiliation{CAS Center for Excellence in Quantum Information and Quantum Physics, University of Science and Technology of China, Hefei 230026, China}

\author{Xi-Wang Luo}\email{luoxw@ustc.edu.cn}
\affiliation{CAS Key Laboratory of Quantum Information, University of Science and Technology of China, Hefei 230026, China}
\affiliation{CAS Center for Excellence in Quantum Information and Quantum Physics, University of Science and Technology of China, Hefei 230026, China}

\author{Chang Liu}
\affiliation{CAS Key Laboratory of Microscale Magnetic Resonance and School of Physical Sciences, University of Science and Technology of China, Hefei 230026, China}
\affiliation{CAS Center for Excellence in Quantum Information and Quantum Physics, University of Science and Technology of China, Hefei 230026, China}

\author{Yue Li}
\affiliation{CAS Key Laboratory of Microscale Magnetic Resonance and School of Physical Sciences, University of Science and Technology of China, Hefei 230026, China}
\affiliation{CAS Center for Excellence in Quantum Information and Quantum Physics, University of Science and Technology of China, Hefei 230026, China}

\author{Mingdong Zhu}
\affiliation{CAS Key Laboratory of Microscale Magnetic Resonance and School of Physical Sciences, University of Science and Technology of China, Hefei 230026, China}
\affiliation{CAS Center for Excellence in Quantum Information and Quantum Physics, University of Science and Technology of China, Hefei 230026, China}

\author{Xi Qin}
\affiliation{CAS Key Laboratory of Microscale Magnetic Resonance and School of Physical Sciences, University of Science and Technology of China, Hefei 230026, China}
\affiliation{CAS Center for Excellence in Quantum Information and Quantum Physics, University of Science and Technology of China, Hefei 230026, China}

\author{Yiheng Lin} \email{yiheng@ustc.edu.cn}
\affiliation{CAS Key Laboratory of Microscale Magnetic Resonance and School of Physical Sciences, University of Science and Technology of China, Hefei 230026, China}
\affiliation{CAS Center for Excellence in Quantum Information and Quantum Physics, University of Science and Technology of China, Hefei 230026, China}

\author{Jiangfeng Du} \email{djf@ustc.edu.cn}
\affiliation{CAS Key Laboratory of Microscale Magnetic Resonance and School of Physical Sciences, University of Science and Technology of China, Hefei 230026, China}
\affiliation{CAS Center for Excellence in Quantum Information and Quantum Physics, University of Science and Technology of China, Hefei 230026, China}

\begin{abstract}
Triply degenerate points (TDPs), which correspond to 
new types of topological semimetals,
can support novel quasiparticles possessing effective integer spins while preserving Fermi statistics. Here by mapping the momentum space to
the parameter space of a three-level system in a trapped ion, we experimentally explore the transitions between different types of TDPs driven by spin-tensor--momentum couplings. 
We observe the phase transitions between TDPs with different topological charges by measuring the Berry flux on a loop surrounding the gap-closing lines, 
and the jump of the Berry flux gives the jump of the topological charge (up to a $2\pi$ factor) across the transitions. 
For the Berry flux measurement, we employ a new method by examining the geometric rotations of both spin vectors and tensors, which lead to a generalized solid angle equal to the Berry flux. The controllability of  multi-level ion offers a versatile platform to study high-spin physics and our work paves the way to explore novel topological phenomena therein. 
\end{abstract}

\date{\today}

\maketitle  

\emph{Introduction.---}Topological states of matter, including topological insulators, superconductors and semimetals, have attracted increasing interest in the past decades~\cite{RMP_Graphene,RMP_TI,RMP_TI_SC}. Recent studies on topological semimetal 
had led to the observation of Weyl~\cite{Weyl_point_wan,Weyl_point_Hirschberger,Weyl_point_Lv,Weyl_point_Xu,Weyl_point_Wang} and Dirac~\cite{Dirac_point_Xiong,Dirac_point_Liu} Fermions in solid-state materials, which possess two- or four-fold degenerate points and support relativistic spin-1/2 quasi-particles. Very recently, the remarkable discovery of triply degenerate points (TDPs)~\cite{TDP_Bradlyn,TDP_Hu,TDP_Fulga,Monopole_Hu,TDP_Lv,TDP_Yang,TDP_Tan,TDP.Yang.2017,TDP.Weng.2016,TDP.Zhu.2016,TDP.Zhong.2017}  in Fermionic systems provides an avenue for exploring new types of quasiparticles possessing integer spins while preserving Fermi
statistics that have no counterparts in quantum field theory.
The TDPs (i.e., three-fold band degeneracies in spin-1 systems) behave like magnetic monopoles in momentum space whose
topological charges $\mathcal{C}$ are
determined by the Berry flux emanating from the
degenerate points. Unlike the spin-1/2 particles, a full characterization of higher spins ($\geq 1$) naturally involves both the spin vectors $\hat{\mathbf{F}}$ and high-rank spin tensors such as $\hat{N}_{ij}=\{\hat{F}_i,\hat{F}_j\}/2-\delta_{ij}\hat{\mathbf{F}}^2 /3$. 
Therefore, an important question is to explore the roles played by spin tensors in driving the phase transition and characterizing the topologies of the TDPs.

Previous studies have predicted that 
spin-tensor momentum couplings 
can induce transitions between TDPs with different monopole charges $\mathcal{C}=\{0,\pm 1,\pm 2\}$~\cite{TDP_Hu,Monopole_Hu}. On the other hand,
the Berry flux and monopole charge cannot be solely
determined by the solid angle of spin vector 
and its covering number on
the Bloch sphere as in spin-1/2 case. In fact, the spin-1 vector  
can go inside the Bloch sphere and the spin tensors 
must also be taken into account to obtain the Berry flux~\cite{h2018non,bharath2018singular,PhysRevB.101.140412,PhysRevA.102.033339}.

Experimentally, TDPs with topological charge $\mathcal{C}=2$ has been observed in various systems, including  solid-state topological semimetal molybdenum phosphide~\cite{TDP_Lv}, phononic crystal~\cite{TDP_Yang}, as well as in the synthetic parameter space of a superconducting qutrit~\cite{TDP_Tan}. 
In contrast to condensed matter systems where the realization of required spin-momentum coupling and the measurement of topological properties would be challenging,  synthetic quantum systems with versatile control (e.g., cold atom~\cite{Science.360.1429,PhysRevLett.127.136802}, superconducting qubit~\cite{Nature.515.241,PhysRevLett.113.050402,PhysRevLett.122.210401,PhysRevLett.126.017702}, nitrogen-vacancy center~\cite{PhysRevLett.117.060503,PhysRevLett.120.120501,PhysRevLett.125.020504,ChinRevLett.34.060302}, trapped ion~\cite{RevModPhys.75.281} systems, etc.) offer powerful tools 
for quantum simulation of
topological phenomena in parameter space.
To date, the topological transitions between TDPs with different monopole charges 
and the crucial roles played by the spin tensors have not been demonstrated experimentally. 

In this paper, by mapping the momentum space to
the parameter space of a trapped ion, we experimentally explore the topological transitions between different types of TDPs and
demonstrate the important roles played by the spin tensors, where the Berry flux is measured through
the generalized solid angle traced out by
the trajectories of both spin vectors and tensors. We simulate a momentum space Hamiltonian 
\begin{equation}
    H=\mathbf{k}\cdot \hat{\mathbf{F}}+\alpha k_z \hat{N}_{zz}+\beta k_x \hat{N}_{xz},
    \label{Eq:Ham_1}
\end{equation} 
which describes the pseudospin-1 particles with a TDP at $\mathbf{k}=0$ carrying topological charges depending on the spin-tensor--momentum coupling strengths
$(\alpha,\beta)$~\cite{SM}. We effectively tune 
$(\alpha,\beta)$ and observe the transitions of the TDPs from $\mathcal{C}=2$ to $\mathcal{C}=1$ and $0$ by measuring the spin vectors and tensors. At the transitions,
we observe sudden jumps of both the spin vectors (represented by arrows) and tensors (represented by ellipsoids) at the corresponding gap-closing momenta. For the transition from $\mathcal{C}=2$ to $\mathcal{C}=1$, 
the Berry flux is solely determined by the spin vectors, whose jump indicates the jump of monopole charge by 1. On the other hand, for 
the transition from $\mathcal{C}=2$ to $\mathcal{C}=0$, we observe the jump of topological charge by measuring the Berry flux on a small loop surrounding gap-closing momenta. We adiabatically drive the system along the small loop and detect
the generalized solid angle traced out by both the spin vector arrow and tensor ellipsoid, leading to a geometric phase equal to the Berry flux. Crucially, the Berry flux contains contributions from both spin vectors and tensors. 

\begin{figure}[t!]
	\begin{center}

	\includegraphics[width=1\columnwidth]{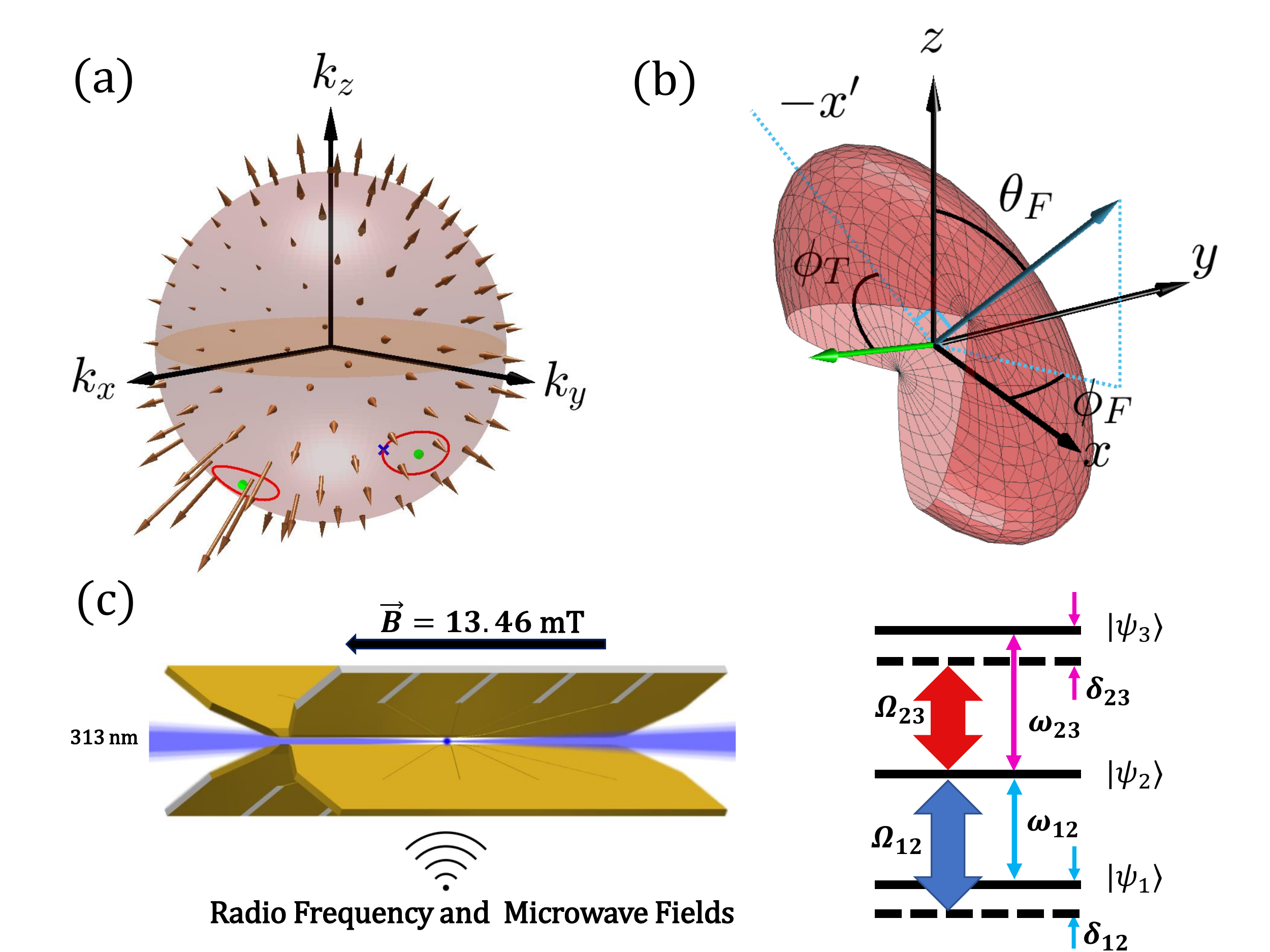}
	\caption{Berry flux distribution, state geometric presentation and experimental setup. (a) Berry flux distribution on momentum sphere with $\alpha=0,~\beta=-1$. Green dots are the gap-closing points $\mathbf{k}_c$ surrounded by red loops. (b) Spin tensor ellipsoid at momentum represented by the blue cross on one of the loops in (a). The longitudinal (transverse) direction  of the ellipsoid is given by the spin-vector arrow in blue (short-axis arrow in green).  Azimuthal angle $\phi_T$ of the tensor ellipsoid is given by the relative angle between green arrow and axis $-x^\prime$, where $x^\prime$ is the rotated axis $x$ with Euler angles $(0,\theta_F,\phi_F)$ as defined in the main text. (c) Illustration of the experimental setup, energy levels and transitions. A three-level trapped $^9\rm{Be}^+$ ion is driven by radio frequency and microwave fields, forming Hamiltonian in Eq.~(\ref{Eq:Ham_exp}).
	}
	\label{fig:flux_expsetup}
	\end{center}
\end{figure}

\emph{Model and experimental setup.---}We consider a three-band spin-1 system with Hamiltonian given by Eq.~(\ref{Eq:Ham_1}). The momentum space can be parameterized by the spherical coordinates $\mathbf{k}=k_0(\sin\theta\cos\phi, \sin\theta\sin\phi,\cos\theta)$, and the TDP appears at $k_0=0$ where all three bands degenerate at zero energy. The bands open gaps for $k_0>0$ with monopole charge
$\mathcal{C}=\frac{1}{2\pi}\oint \mathbf{\Omega_{k}}\cdot d \mathbf{\mathcal{S}} $ given by the total Berry flux on the sphere $\mathbf{\mathcal{S}}$ surrounding the TDP (see Fig.~\ref{fig:flux_expsetup}a), where $\mathbf{\Omega_k}=\nabla_\mathbf{k}\times \mathbf{A_k}$ and $\mathbf{A_k}=\langle \Psi(\mathbf{k})|i\nabla_\mathbf{k} | \Psi(\mathbf{k})\rangle $ are the Berry curvature and connection respectively, and $|\Psi(\mathbf{k})\rangle$ is the eigenstate for the lowest band. 
The band gaps close along certain lines (i.e., gap-closing points $\mathbf{k}_c$ are $\{\theta_c,\phi_c,\forall k_0\}$) as we change ($\alpha, \beta$) across the phase transitions where the topological charge of the TDP changes. 

A direct measurement of the topological phase transition requires measuring
the change of the total Berry flux on the surface enclosing the TDP monopole. A spin-1 quantum state is determined by the mean values of both spin vectors and tensors, which are represented by an arrow and
an ellipsoid~\cite{h2018non,bharath2018singular,PhysRevB.101.140412,PhysRevA.102.033339}, respectively. The orientation and size of the ellipsoid are given by the eigenvectors and the square root of the eigenvalues of the tensor matrix $T_{ij}=\braket{\hat{N}_{ij}}-\braket{\hat{F}_i}\braket{\hat{F}_j}+2\delta_{ij}/3$, as shown in Fig.~1b. We can measure the Berry flux $\gamma=\int \mathbf{\Omega_{k}}\cdot d \mathbf{S}_{\mathcal{L}}=\oint_{\mathcal{L}} \mathbf{A_{k}}\cdot d\mathbf{k}$ through an area $\mathbf{S}_{\mathcal{L}}$ surrounded by a loop $\mathcal{L}$ on the sphere, which can be 
obtained from the geometric rotations of both the arrow and ellipsoid 
$\gamma=\gamma_F+\gamma_T$ where
\begin{eqnarray}
\gamma_F\equiv\oint_{\mathcal{L}} F\cos\theta_F d\phi_F \text{ and }
\gamma_T\equiv\oint_{\mathcal{L}} F d\phi_T
\label{Eq:Berryflux}
\end{eqnarray}
are the generalized solid angles for the spin vector and tensor~\cite{PhysRevA.102.033339,SM}, respectively.
Here $F$ and $(\theta_F,\phi_F)$ are the length and spherical angles of spin vector $\braket{\hat{\mathbf{F}}}$, $\phi_T$ is the relative rotation angle of the spin-tensor ellipsoid with respect to the spin vector (see Fig.~\ref{fig:flux_expsetup}b).
Therefore,
the monopole charges and topological phase transitions can be characterized by the rotations of the spin-vector arrows and spin-tensor ellipsoids 
which can be directly detected in experiments.

To simulate such a spin-1 system we map the momentum space to the parameter space of a trapped ion, whose three coupled internal states form a pseudospin-1 system.
We trap a single $^9\rm{Be}^+$ ion in a linear Paul trap \cite{Li_2021} with ambient magnetic field of 13.46 mT (see Fig.~\ref{fig:flux_expsetup}c). Three states, denoted as ${|\psi_1\rangle,|\psi_2\rangle,|\psi_3\rangle}$ respectively, in the ground manifold $2s~^2 S_{1/2}$ are utilized (see \cite{SM} for detailed definitions), which form a spin-1 system, as shown in Fig.~\ref{fig:flux_expsetup}c. Resonant transition frequencies between states $|\psi_i\rangle$ and $|\psi_j\rangle$ are denoted as $\omega_{i j}$, where $\omega_{1 2}=2\pi\times118.966\ \rm{MHz}$ and $\omega_{2 3}=2\pi\times991.570\  \rm{MHz}$. To drive these transitions, we apply impedance matched antennas \cite{Li_2021} connected to power-amplified signal sources to induce radio-frequency (RF) and microwave fields to the ion, respectively, where the former is sourced by an arbitrary-wave-generator (AWG) and the latter is sourced by a separate AWG, frequency-mixed with a high frequency microwave source of approximately 1 GHz. Such a configuration combining the RF and microwave transitions enables us to directly drive each transition within the ground state manifold satisfying the selection rules, and thus would be readily scalable to include more levels, particularly for demonstrations where tailored connectivity are required \cite{Monopole_Ray}. By programming the AWGs with desired waveform, we apply time-dependent drives with Rabi rate $\Omega_{i j}$, detuning $\delta_{i j}$, and phase $\phi_{i j}$, as depicted in Fig.~\ref{fig:flux_expsetup}. Thus in a rotating wave approximation, we obtain the desired Hamiltonian in the
$\{|\psi_i\rangle\}$ basis
\begin{equation} H = 
\begin{pmatrix}
\delta_{12} & \Omega_{12}e^{i \phi_{12}} & 0\\
\Omega_{12}e^{-i \phi_{12}} & 0 & \Omega_{23}e^{i \phi_{23}} \\
0 & \Omega_{23}e^{-i \phi_{23}} & \delta_{23} \\
\end{pmatrix},\label{Eq:Ham_exp}\end{equation}
which can further be expressed by the spin-1 monopole Hamiltonian Eq.~(\ref{Eq:Ham_1}) (up to a constant), with detunings, coupling amplitudes and phases
given by $\delta_{12}=(\alpha+1)k_z=k_0(\alpha+1)\cos
\theta$, $\delta_{23}=(\alpha-1)k_z=k_0(\alpha-1)\cos
\theta$, $\Omega_{12}e^{i \phi_{12}}=(1+\beta/2)k_x-ik_y=\frac{k_0\sin\theta}{\sqrt{2}} e^{-i\phi}+\beta\frac{k_0 \sin\theta}{2\sqrt{2}} \cos\phi$ and
$\Omega_{23}e^{i \phi_{23}}=(1-\beta/2)k_x-ik_y=\frac{k_0\sin\theta}{\sqrt{2}} e^{-i\phi}-\beta\frac{k_0 \sin\theta}{2\sqrt{2}} \cos\phi$. Here, $k_0$ only 
modifies the magnitudes of the energy bands without affecting the eigenstates, therefore, we focus our discussions on a sphere with fixed $k_0$.

The experiment begins with a series of controlled 313 nm laser beam pulses (Fig.~\ref{fig:flux_expsetup}c) to Doppler cool the ion motion and initialize it to $|\psi_2\rangle$ to further couple to the other states. We then apply a sequence of resonant RF and microwave pulses to prepare the ion to the ground state of the Hamiltonian for given parameters $\{\alpha,\beta,\theta,\phi\}$, where the amplitudes and durations of the control pulses can be calculated via diagonalizing the Hamiltonian. To measure the Berry flux within a loop, we subsequently apply an adiabatic ramp of the parameters $(\theta,\phi)$ along the loop of interest on the sphere with fixed $\alpha$ and $\beta$. We stop the evolution at various points on the loop, and measure the observables $\langle \hat{F}_i \rangle$, $\langle \hat{N}_{ij}\rangle$ for $\{i,j\}=\{x,y,z\}$. In separate experimental trials we map each of the eigenstate of the observables to $|\psi_2\rangle$ by applying analytically tailored resonant pulse sequences \cite{SM}, while the other states are mapped to far away states within the same ground manifold followed by a resonant fluorescence detection for 400 $\rm{\mu s}$. We typically collect on average 25 counts for $|\psi_2\rangle$ and up to average of 1.6 counts for the other states. By setting a threshold of 6 counts and repeating the experiment with 500 trials, we distinguish the populations and further obtain the desired expectation values \cite{SM}. 

\begin{figure}[t!]
	\begin{center}
		\includegraphics[width=1\columnwidth]{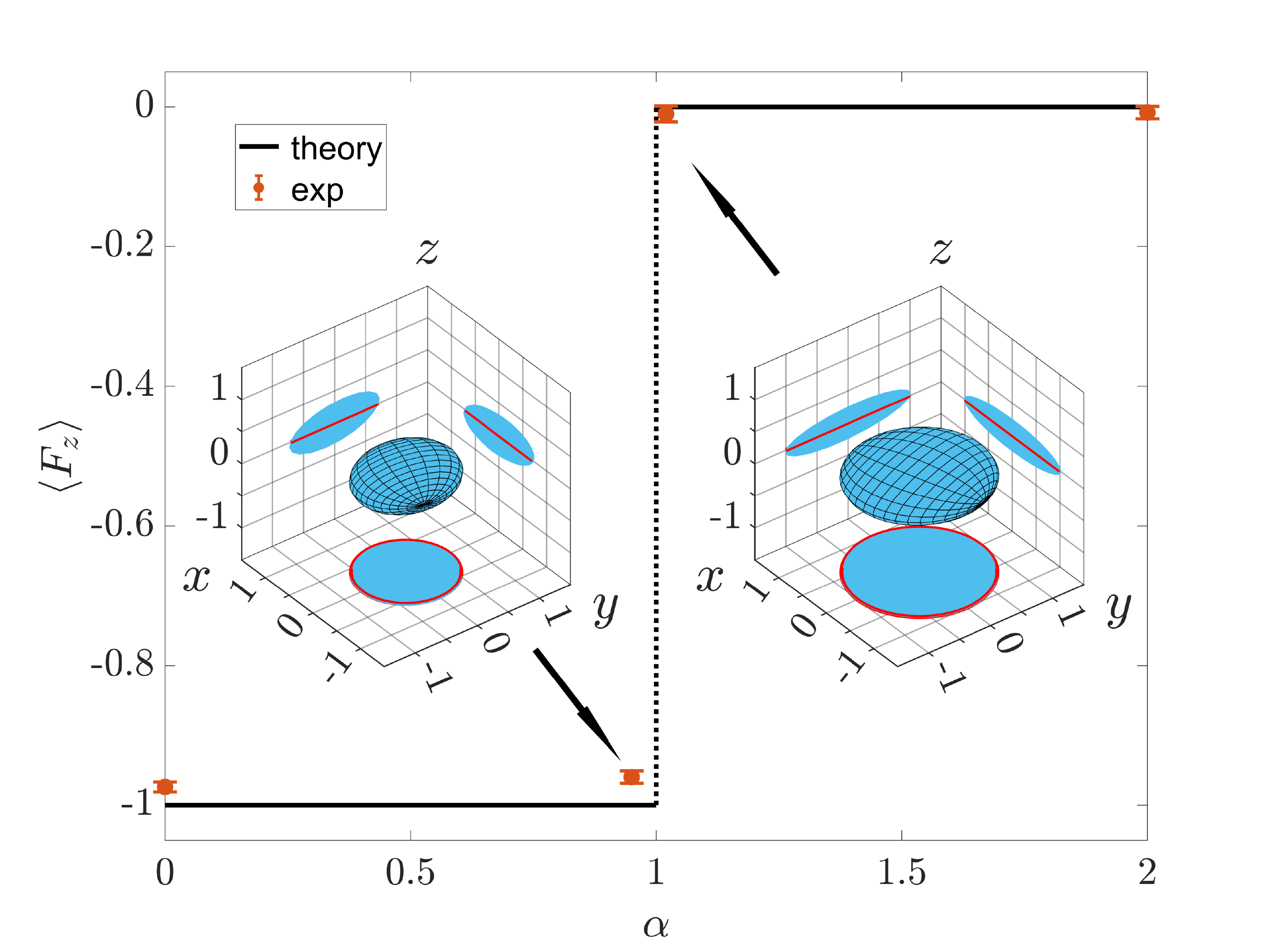}
        \caption{Phase transition characterized by the jump of spin vector and tensor. Errorbars correspond to one standard deviation.
        The insets describe tensor ellipsoids and their projection on $x,~y$ and $z$ plane at $\alpha=0.95$ (left with $\mathcal{C}=2$) and $\alpha=1.05$ (right with $\mathcal{C}=1$), with $\beta=0$. The red circles and lines are the projections of the theoretical tensor ellipsoids. The experimental imperfection leads to a finite axis length of the ellipsoid along $z$ direction $\sim0.3$, corresponding to a bias of $0.3^2\sim 10\%$ in measuring $\hat{N}_{ij}$. 
		}
		\label{fig:beta}
	\end{center}
\end{figure} 

\emph{Observation of the topological phase transitions.---}We first set $\beta=0$ and consider the transition from $\mathcal{C}=2$ to
$\mathcal{C}=1$ by increasing $\alpha$.
The band gaps close at the north pole (i.e., $\theta_c=0$) on the momentum sphere as $\alpha$ changes across $\alpha_c=1$. 
We measure the corresponding spin vectors $\braket{\hat{F}_i}$ and tensors
$\braket{\hat{N}_{ij}}$ of the ground state at $\theta=0$ for different $\alpha$. As depicted in Fig. \ref{fig:beta}, the measured value of $\braket{\hat{F}_z}$ for $\alpha<1$ is approximately equal to $-1$ but dramatically jumps to approximately 0 when $\alpha>1$ ($\braket{\hat{F}_{x}}$ and $\braket{\hat{F}_{y}}$ are always approximately equal to 0), indicating a phase transition. For a spin-1 system, the states are characterized by the expected values of both spin vectors $\braket{\hat{F}_i}$ and tensors $\braket{\hat{N}_{ij}}$.
Geometrically, the spin vector corresponds to an arrow, while the spin tensor can be described by an ellipsoid. As also depicted in Fig. \ref{fig:beta}, we observe a dramatic change of ellipsoid around the phase transition $\alpha=1$. Moreover, we observe the
spin vortex at the north pole for $\alpha>1$ \cite{SM}, which also signals the transition of the monopole charge~\cite{Monopole_Hu}. To illustrate the jump of monopole charge, 
we examine latitude loops on the momentum sphere since the Hamiltonian has cylindrical symmetry with respect to $z$ axis, leading to $\gamma_F=-2\pi \cdot\braket{\hat{F}_z}$ \cite{SM}. The tensor has no contribution to the Berry flux where $\phi_T$ is always 0 in the case $\beta=0$.
From Fig.~2, we observe $\braket{\hat{F}_z}$ approximately  changes by 1 at the north pole ${\theta=0}$, matching with the expected  $2\pi$ change of the Berry flux, and thus the monopole charge $\mathcal{C}=\braket{\hat{F}_z}|^{\theta=\pi}_{\theta=0}$ 
changes by 1.

\begin{figure}[t!]
	\begin{center}
		\includegraphics[width=0.9\columnwidth]{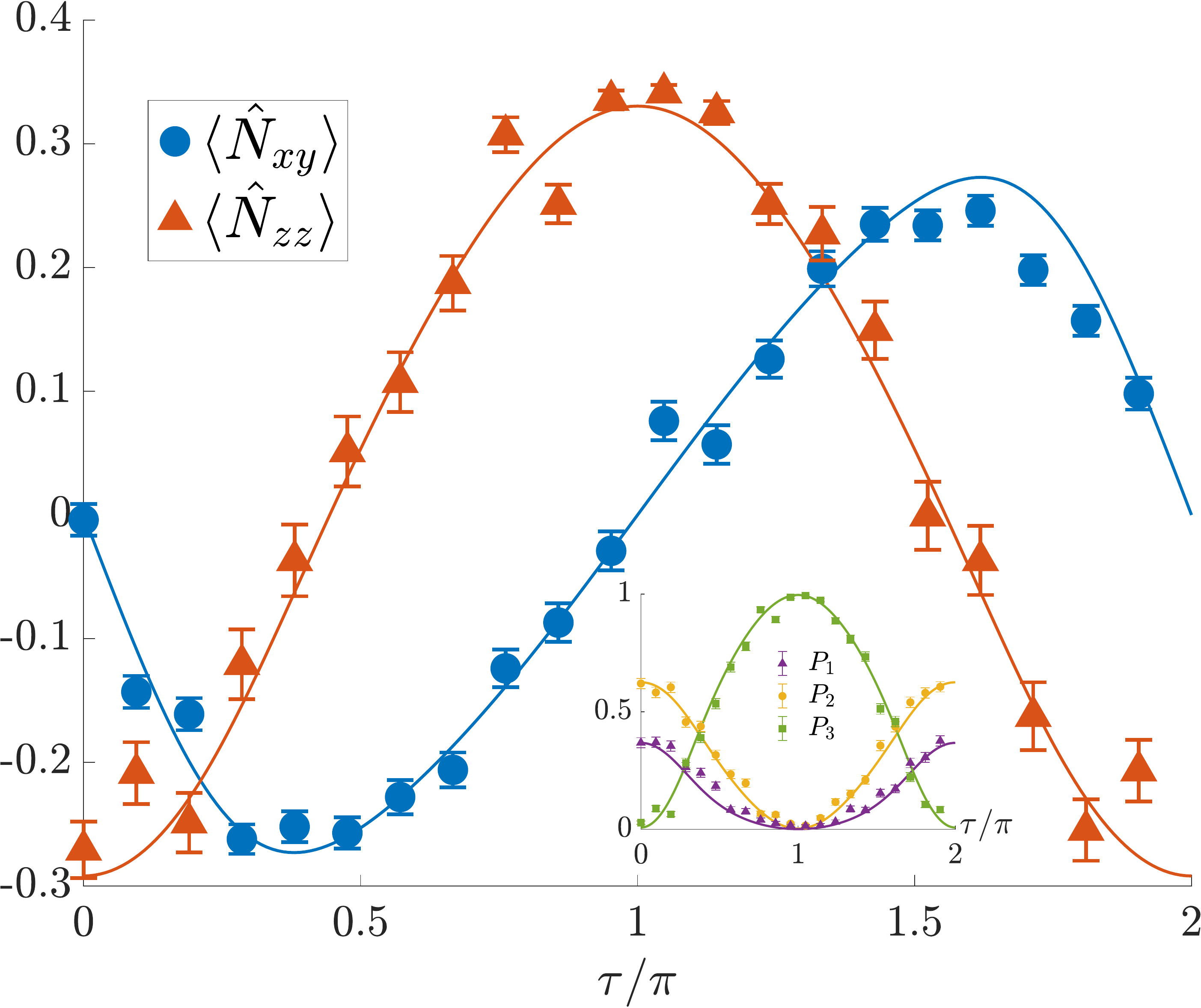}
        \caption{Measured spin tensors at different position $\tau$ along the loop $\mathcal{L}$ with $\alpha=0,~\beta=-1.9$. Blue circles and red triangles are $\langle \hat{N}_{xy}\rangle$ and $\langle \hat{N}_{zz}\rangle$ respectively (more data can be found in \cite{SM}). In the inset, green squares, yellow circles and purple triangles show the populations $P_{i=\{1,2,3\}}$ respectively for the
        eigenstates of ${\hat{N}_{zz}}$, leading to $\langle \hat{N}_{zz}\rangle=\sum_iP_i\epsilon_i$ with $\epsilon_{i=\{1,2,3\}}=\{1/3,-2/3,1/3\}$ the eigenvalues of $\hat{N}_{zz}$. 
        Errorbars correspond to one standard deviation. Solid lines represent the corresponding numerical simulations.
		}
		\label{fig:Fcompon}
	\end{center}
\end{figure} 
For a general spin-1 model, both the vectors and tensors should contribute to the Berry flux. 
We notice that at the vicinity of the phase transition, the sudden change of the monopole charge must be given by the sudden change of the Berry flux near the non-analytical point (i.e., the gap closing point). Therefore, measuring the Berry flux near the gap closing point can be used to probe the topological phase transition directly. To show this, we examine the topological phase transition from $\mathcal{C}=2$ to $\mathcal{C}=0$ and set $\alpha=0,~\beta\neq0$ for
a different spin-tensor--momentum coupling.
The band gap closes at $(\theta_c,\phi_c)=(3\pi/4,0)$ and $(3\pi/4,\pi)$ on the momentum sphere across the phase transition point $\beta_c=-2$, and we expect jumps of both spin vectors and tensors \cite{SM}.

\begin{figure}[t!]
	\begin{center}
		\includegraphics[width=1\columnwidth]{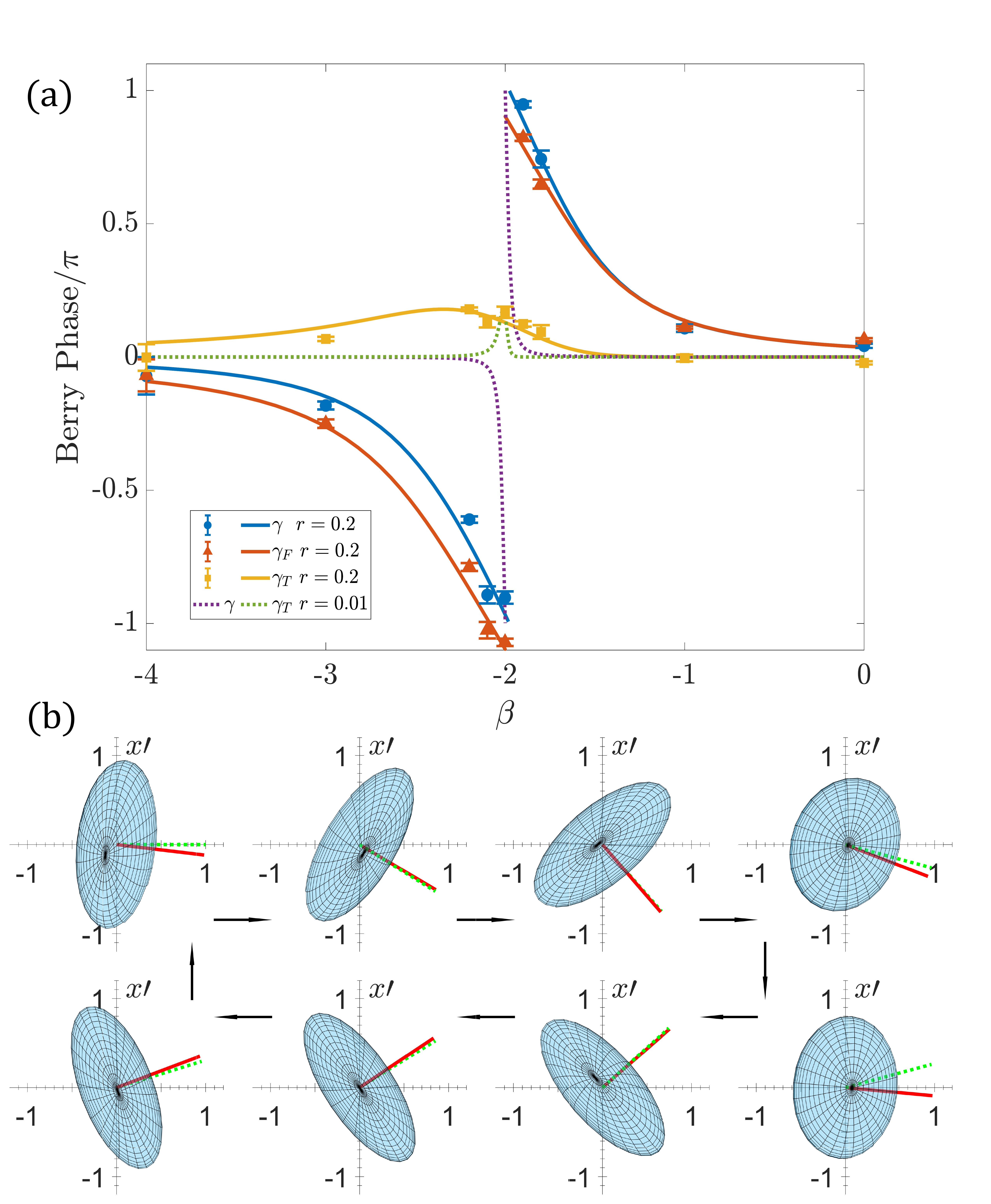}
        \caption{Berry phase and rotation of corresponding ellipsoids. (a) Berry flux $\gamma$ through the loop $\mathcal{L}$ versus $\beta$. Blue circles, red triangles and yellow squares are the experimental data of Berry flux $\gamma$, vector contribution $\gamma_F$ and tensor contribution $\gamma_T$ respectively for the loop with $r=0.2$, the solid lines are the corresponding numerical simulations in the adiabatic limit. Purple and green dashed lines are the numerical simulations of $\gamma$ and $\gamma_T$ with $r=0.01$ respectively. Errorbars correspond to one standard deviation. 
        (b) Measurements of the tensor ellipsoids along the adiabatic loop at $\beta=-2.2$, with $\tau=\{0,0.19,   0.48,0.95,1.05,1.71,1.81,1.90\}\pi$ starting from the top left along the direction indicated by black arrows. The red solid (green dashed) lines represent the direction of the short axis of the ellipsoid from experimental data (numerical simulations), showing the evolution of $\phi_T$ and the rotation with respect to $\braket{\hat{\mathbf{F}}}$.
		}
		\label{fig:beta_sum}
	\end{center}
\end{figure} 

To measure the change of monople charge, we consider a small loop on the momentum sphere surrounding the gap-closing point. After the ground state preparation of the initial Hamiltonian, we evolve the state by subsequently applying an adiabatic ramp of the parameters along the loop $\mathcal{L}$: $\theta=\frac{3\pi}{4}-\frac{3}{4}r \rm{cos}\tau$ and $\phi=\pi-\sqrt{3}r \rm{sin}\tau$ with a nearly uniform gap, where we ramp $\tau$ from 0 to $2\pi$ with a constant rate. By programming separate channels of the AWG, we generate the desired RF and microwave fields as
$f_{ij}(t)=A_{ij}(t)\cos{[\omega_{ij}t+\int_0^t\delta_{ij}(t^\prime)dt^\prime+\phi_{ij}(t)]}$ \cite{SM}, where $t=T\tau/(2\pi)$ is the duration of the waveform with $T$ the maximum ramp time of 1 ms to enclose a loop, $A_{ij}$ is the experimental amplitude of waveform corresponding to Rabi frequency $\Omega_{ij}$.
We choose $r=0.2$ and the ramp rate is separately checked via a numerical simulation to ensure a required level of adiabaticity and coherence \cite{SM}. We observe a number of $\langle \hat{F}_i \rangle$, $\langle \hat{N}_{i j}\rangle$ at various $\tau$ (see Fig.~\ref{fig:Fcompon} for $\beta=-1.9$ as an example) by measuring the eigenstate populations of these observables. From these values, we can obtain the spin-vector length $F=|\langle \hat{\mathbf{F}} \rangle|$, the spherical angles ${\theta_F,\phi_F}$ denoting the relative angle from the $z$ direction and the projection angle to the $x$-$y$ plane, as shown in Fig. \ref{fig:flux_expsetup}b. Importantly, we can obtain the relative rotation angle $\phi_T$ of the spin-tensor-ellipsoid with respect to the spin vector \cite{SM}. With these information, we finally arrive to the Berry phase 
$\gamma=\gamma_F+\gamma_T$.
By repeatedly measuring the Berry flux over a selection of $\beta$, we observe the Berry flux changes {from 0 at $|\beta+2| \gg 0$ to approximately $\pm\pi$  at $|\beta+2|=0$}, with a sharp transition by $2\pi$ at $\beta = -2$ (see Fig.~\ref{fig:beta_sum}a). Similarly, one could apply measurement of the Berry flux on the loop around 
$\theta_c=3\pi/4,\phi_c=0$, and the Berry flux should also change by $2\pi$. As a result, the monople charge must change by 2 across the phase transition at $\beta_c=-2$, i.e., $\mathcal{C}$ changes from 2 to 0.
For such a phase transition, both the spin vectors and tensors contribute to the Berry flux. 
We plot and observe relative rotations of the tensor ellipsoid with respect to the spin vector along the loop $\mathcal{L}$ with $\beta=-2.2$, as illustrated in Fig.~\ref{fig:beta_sum}b, and the direction of tensor ellipsoids are more sensitive to experimental noises when the two transverse axes have similar length. We find
$\phi_T$ undergoes a sine-like oscillation along the loop while $F$ undergoes a cosine-like oscillation. 
Such rotation gives non-trivial spin-tensor contribution $\oint_{\mathcal{L}} F d\phi_T\simeq0.18\pi$ for the Berry flux around the phase transition. 

Ideally, we should consider an infinitely small loop $r\approx0$ to obtain a very sharp transition exactly at $\beta=-2$, however such an evolution requires infinitely slow ramp rate and measurement resolution, thus not feasible in practice. Nevertheless, a finite size loop with $r=0.2$ is good enough to show the phase transition. This is because, 
for a finite but small loop, 
the Berry flux is also small unless there is a nonanalytical gap-closing point within the loop, so we can restrict the Berry flux to $[-\pi,\pi]$, and the jump from $\pm\pi$ to $\mp\pi$ gives the critical phase transition point. 
Such a jump for $r=0.2$ can be seen around $\beta=-1.98$ in the numerical simulation, away from which, the Berry flux changes smoothly, as shown in Fig.~\ref{fig:beta_sum}a. 

\emph{Conclusion.---}In summary, we experimentally explore the momentum-space spin-1 Hamiltonian and observe the tensor-driven transitions between different types of TDPs with a trapped ion. 
By examining the
vector arrow and tensor ellipsoid properties around the gap-closing points, we experimentally observe the
transitions between different monopole charges of the TDP. Our work
demonstrates the feasibility to measure Berry flux of high-spin systems
based on the generalized
solid angle traced out by the spin moments (vectors and tensors), which paves the way for exploring topological phenomena directly from the geometric rotations of the spin moments in such systems.
Moreover, our study can be generalized to explore topological phenomena for even higher spins (e.g., higher-fold degenerate points~\cite{PhysRevB.93.045113}), since our setup may be readily scaled to more levels within the ground states of $~^9\rm{Be}^+$ ion, and can be extended to more ions with multiple levels therein \cite{qudit_Low,wang_qudit_2020,qudit_Ringbauer}.

\begin{acknowledgments}
\emph{Acknowledgments.---}We acknowledge support from the National Natural Science Foundation of China (grant number 92165206, 11974330),  the Chinese Academy of Sciences (Grants No. XDC07000000), Anhui Initiative in Quantum Information Technologies (Grant No. AHY050000), the USTC start-up funding, and the Fundamental Research Funds for the Central Universities. \\
\end{acknowledgments}

\section{Supplemental Material}
\subsection{Theoretical Details}
\begin{figure} 
    \centering
    \includegraphics [width=1\columnwidth]{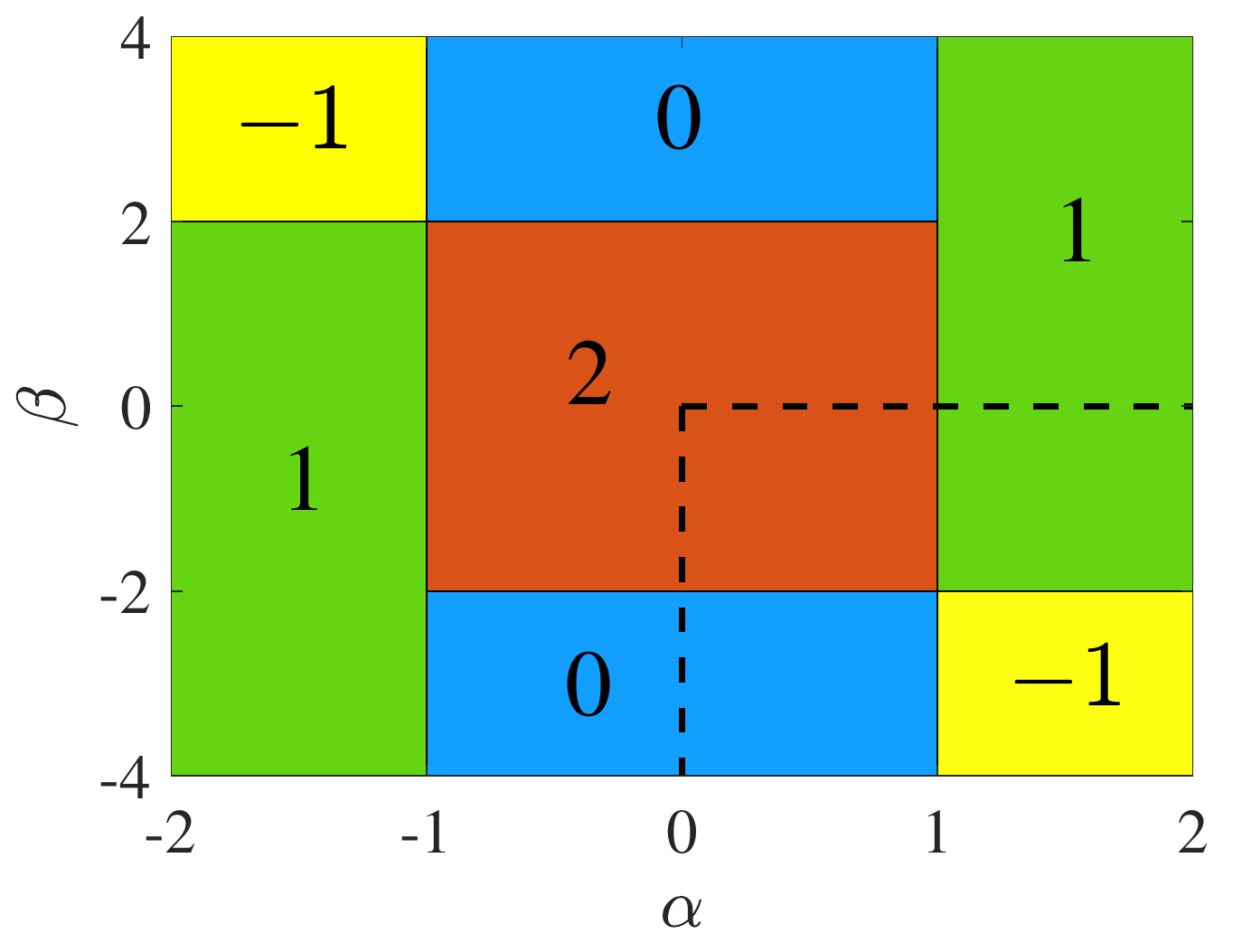}
    \caption{Phase diagram in the $\alpha$-$\beta$ plane. The numbers indicate the monopole charges in each phase.}
    \label{fig:phase_diag}
\end{figure}

\emph{Model and phase diagram.} We simulate a momentum space Hamiltonian 
\begin{equation}
    H=\mathbf{k}\cdot \hat{\mathbf{F}}+\alpha k_z \hat{N}_{zz}+\beta k_x \hat{N}_{xz},
    \label{Eq:Ham_1}
\end{equation} 
which describes the pseudospin-1 particles with a TDP at $\mathbf{k}=0$.
The charge of the TDP is defined as 
$\mathcal{C}=\frac{1}{2\pi}\oint \mathbf{\Omega_{k}}\cdot d \mathbf{\mathcal{S}} $, which
is given by the total ground-state Berry flux on the sphere $\mathbf{\mathcal{S}}$ surrounding the TDP.
The phase diagram in the $\alpha$-$\beta$ plane are shown in Fig.~\ref{fig:phase_diag}, and the monopole 
charge of the TDP can take values of $|\mathcal{C}|=2,1,0$. In this work, we have focused on the transition
from $\mathcal{C}=2$ to $\mathcal{C}=1$ and $0$ along the dashed lines shown in Fig.~\ref{fig:phase_diag}.

\emph{Geometric representation of the quantum state and Berry flux.}
For high spins ($\geq 1$), the spin
moments contain both spin vectors and spin tensors.
Unlike spin-1/2 system, the quantum state is not uniquely represented by
the mean value of spin vector $\langle \hat{\mathbf{F}}\rangle $ which 
is not confined to the surface of the Bloch sphere, and could be
anywhere on or inside the Bloch sphere. Instead, the spin-1 quantum state
is uniquely represented by the combination of the spin vector $\langle
\hat{\mathbf{F}}\rangle $ and rank-2 spin tensors $\braket{\hat{N}_{ij}}$.
Geometrically, the vector is represented by an arrow,
while the tensors is represented by an ellipsoid, whose axes are
determined by the eigenvectors (which give axis directions) and the
square root of the eigenvalues (which give axis lengths) of the rank-2 matrix 
$T_{ij}=\braket{\hat{N}_{ij}}-\braket{\hat{F}_{i}}\braket{\hat{F}_{j}}+2\delta_{ij}/3$.
This rank-2 tensor is also known as the covariance matrix which corresponds to
the fluctuation of the spin vector. For a given vector arrow $\langle
\hat{\mathbf{F}}\rangle $ (with length $F=|\langle
\hat{\mathbf{F}}\rangle|$ and direction angles $\phi _{F},\theta _{F}$), 
the size of the spin-tensor-ellipsoid is also fixed with three axis lengths $\sqrt{1-F^{2}}$%
, $\sqrt{\frac{1\pm \sqrt{1-F^{2}}}{2}}$. Moreover, the axis with length $%
\sqrt{1-F^{2}}$ has the same direction with $\langle\hat{\mathbf{F}}\rangle $,
and the direction of the other two axes with length
$\sqrt{\frac{1\pm \sqrt{1-F^{2}}}{2}}$%
is determined by an azimuthal angle $\phi _{T}$, which fixes the orientation of the ellipsoid~\cite{h2018non,bharath2018singular,PhysRevB.101.140412,PhysRevA.102.033339}.
Therefore, an arbitrary spin-1 quantum
state $|\Psi \rangle $ can be characterized by four parameters $F,\phi
_{F},\theta _{F},\phi _{T}$. In particular, we have
\begin{equation}
|\Psi (F,\phi _{F},\theta _{F},\phi _{T})\rangle =D(\phi _{F},\theta _{F},\phi _{T})\left[
\begin{array}{c}
\sqrt{\frac{1+F}{2}} \\
0 \\
\sqrt{\frac{1-F}{2}}%
\end{array}%
\right]   \label{eq:state_Tensor_S}
\end{equation}
with $D(\phi _{F},\theta _{F},\phi _{T})=e^{-i\hat{F}_{z}\phi
_{F}}e^{-i\hat{F}_{y}\theta _{F}}e^{-i\hat{F}_{z}\phi _{T}}$.
Notice that the state $[\sqrt{\frac{1+F}{2}},0,\sqrt{\frac{1-F}{2}}]^T$
corresponding to a spin vector arrow pointing to the north pole of the Bloch sphere
$\theta_F=0$, and the short transverse axis of the tensor ellipsoid 
is along $x$ direction corresponding to $\phi_T=0$.

Consider the $\tau $-dependent Hamiltonian $H(\tau )$. For an
adiabatic loop in the parameter space $\tau \in \lbrack \tau _{\text{i}%
},\tau _{\text{f}}]$ with the Hamiltonian satisfying $H(\tau _{\text{i}%
})=H(\tau _{\text{f}})$, the corresponding Berry phase of a given gapped
eigenstate is
\begin{eqnarray}
\gamma &=&\oint_{\mathcal{L}} \mathbf{A_{k}}(\tau)\cdot d\mathbf{k}(\tau) \nonumber \\
&=&i\oint d\tau \langle \Psi (\tau )|\partial _{\tau }|\Psi (\tau
)\rangle,  \label{eq:Berry_S}
\end{eqnarray}
where $|\Psi (\tau )\rangle $ is the
eigenstate of $H(\tau )$ with $|\Psi (\tau _{\text{f}%
})\rangle =|\Psi (\tau _{\text{i}})\rangle $. 
Using Eqs.~(\ref{eq:state_Tensor_S}) and (\ref{eq:Berry_S}), 
we obtain $\gamma =\gamma _{F}+\gamma _{T}$ with
\begin{eqnarray}
\gamma _{F} &\equiv &[\phi _{F}(\tau _{\text{i}})-\phi _{F}(\tau _{\text{f}%
})]+\oint F\cos (\theta _{F})d\phi _{F},  \nonumber \\
\gamma _{T} &\equiv &[\phi _{T}(\tau _{\text{i}})-\phi _{T}(\tau _{\text{f}%
})]+\oint Fd\phi _{T}.
\label{Eq:Berry}
\end{eqnarray}%
For cases studied in this paper,
we can set $\phi _{F}(\tau _{\text{i}})=\phi _{F}(\tau _{\text{f}})$ and 
$\phi _{T}(\tau _{\text{i}})=\phi _{T}(\tau _{\text{f}})$
with proper choice of initial and final angles ($\phi _{F}$ and $\phi _{T}$). So we arrive
\begin{eqnarray}
\gamma _{F} &= &\oint F\cos (\theta _{F})d\phi _{F},  \nonumber \\
\gamma _{T} &=& \oint Fd\phi _{T}.
\label{Eq:Berry2}
\end{eqnarray}%
From the definition, we see that $\gamma _{F}$ ($\gamma _{T}$) corresponds
to the rotation of the spin vector arrow (tensor ellipsoid). 

\begin{figure} 
    \centering
    \includegraphics [width=1\columnwidth]{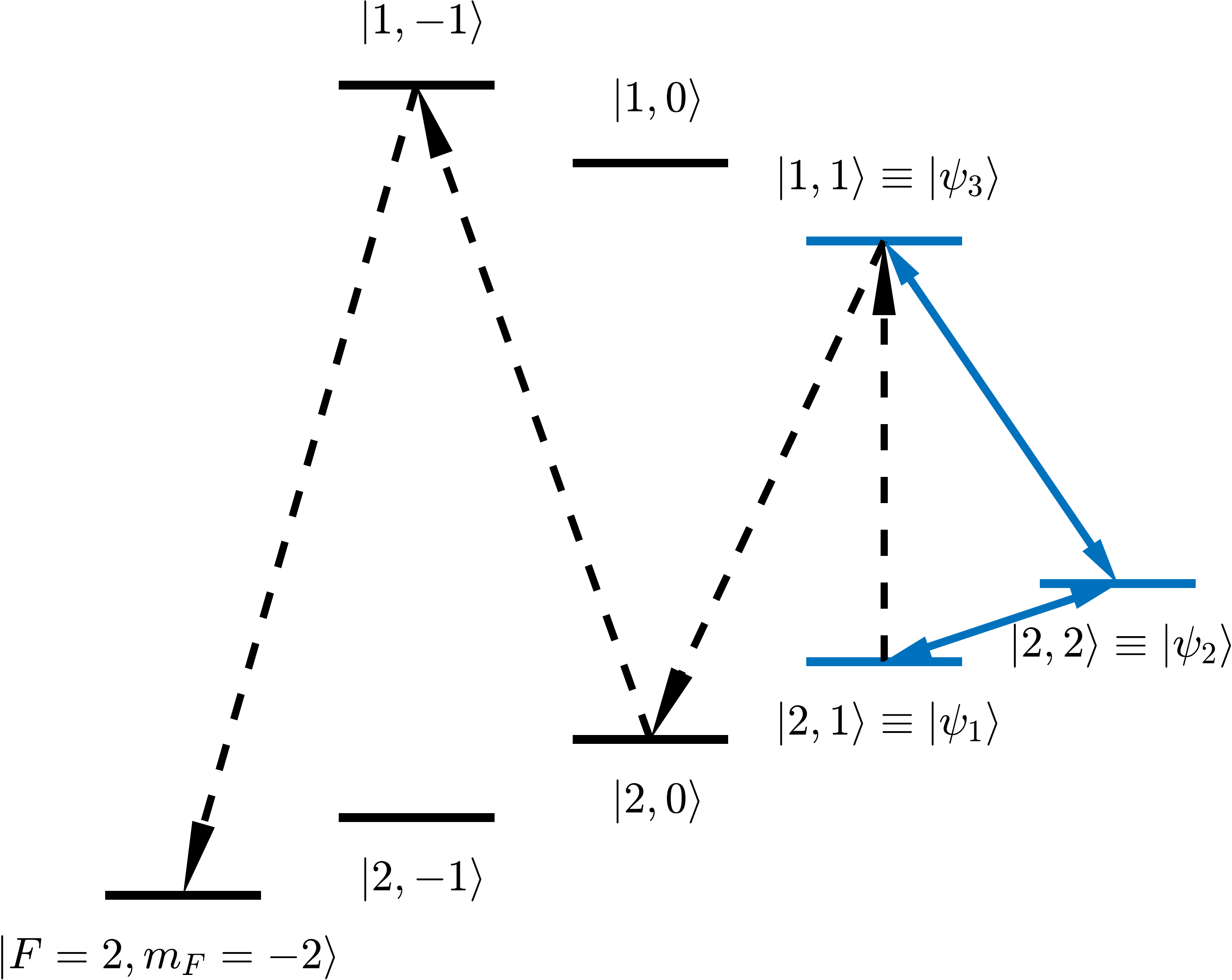}
    \caption{The energy levels of the ground state of $^9\rm{Be}^+$. The dashed black arrows represent the shelving process and the solid blue arrows represent the radio frequency and microwave drives.}
    \label{fig:energy_level}
\end{figure}

\begin{figure} 
    \centering
    \includegraphics [width=1\columnwidth]{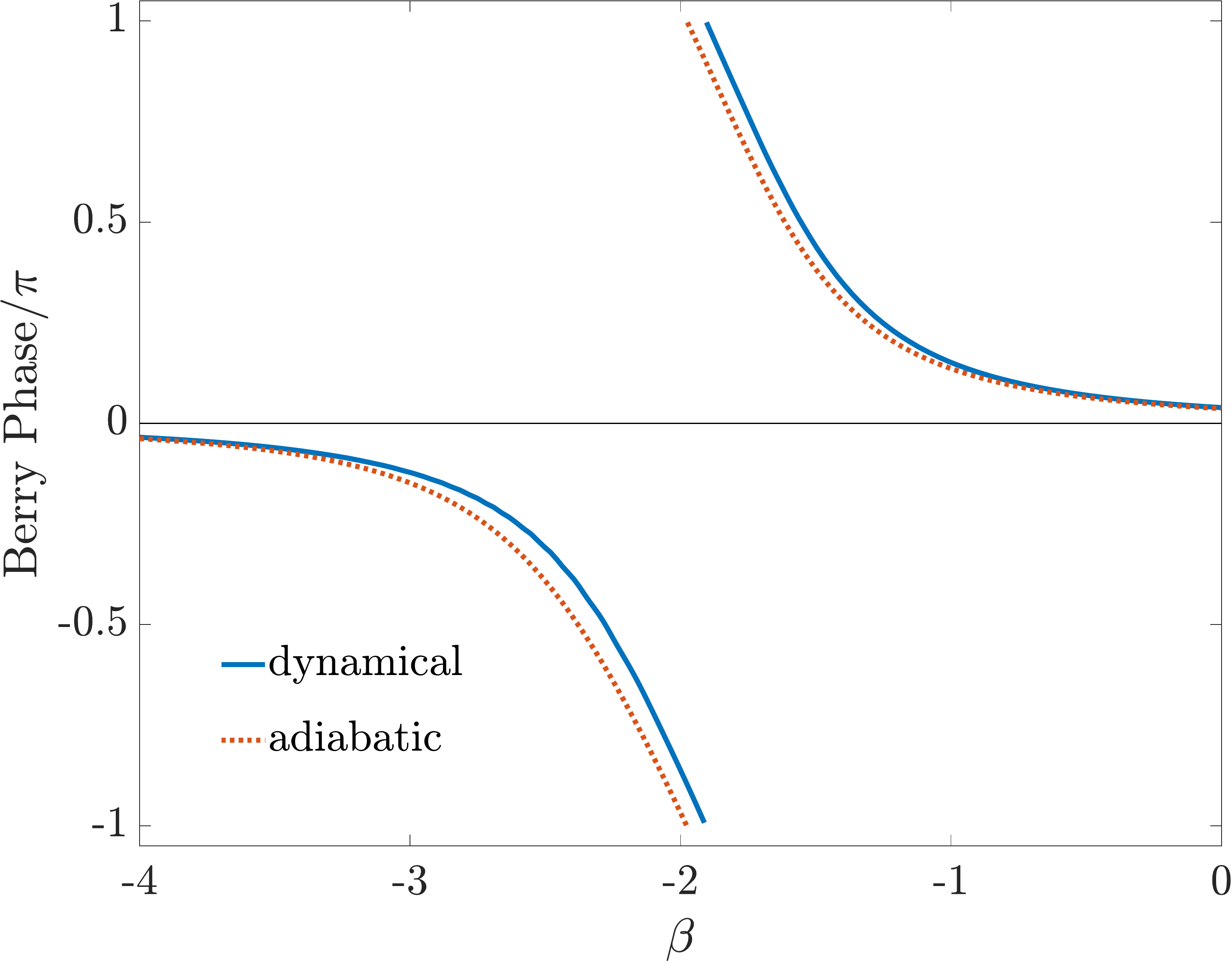}
    \caption{Numerical comparison for Berry phase between in the perfect adiabatic limit and with total ramp time of 1 ms, showing no significant deviations. We set $\pi/k_0=10.67$ us, $\alpha = 0$, $r=0.2$. }
    \label{fig:adiacon}
\end{figure}

For the case with $\beta=0,~\alpha\neq0$, we always have $\phi_T=0$, so the Berry flux is given by
the vector rotation only. Due to the cylindrical symmetry, $\braket{\hat{F}_z}$ is independent from
$\phi_F$, and we have $\gamma=\braket{\hat{F}_z} \oint d\phi_F$ and $\phi_F=0$ or $\phi_F=\pm \phi$ for latitude loops on the momentum sphere.
Here the eigenstate satisfies $\phi_F=-\phi$, so  $\gamma=-2\pi\braket{\hat{F}_z}$. The topological charge, given by the Berry flux difference between latitude loops at the south and north poles, is $\mathcal{C}=\braket{\hat{F}_z}|^{\theta=\pi}_{\theta=0}$. 
We want to mention that
the Berry curvature on the momentum sphere $\Omega_{\theta\phi}=\rm{Im}[\braket{\partial_\theta \Psi|\partial_\phi \Psi}-\braket{\partial_\phi \Psi|\partial_\theta \Psi}]$, given by the derivative of $\gamma$, reads
$\Omega_{\theta\phi}=\partial_\theta \braket{\hat{F}_z}$, and the charge
$\mathcal{C}=\frac{1}{2\pi}\int\Omega_{\theta\phi} d\theta  d\phi $.

For the case with $\beta\neq0,~\alpha=0$, the Berry flux and curvature take complicated values, with contributions from both the spin vector and tensors. 
Experimentally, $\phi_T$ can be extracted by first rotating the vector arrow and tensor ellipsoid though
Euler angles $(-\phi_F,-\theta_F,0)$, then $\phi_T$ is given by the azimuthal angle of the
short transverse axis of the ellipsoid.


\subsection{Experimental Details}


\emph{Experimental setup and data processing.} For $^9\rm{Be}^+$ ion, the hyperfine structure of ground state can be represented by total angular momentum $F$ and $m_F$, as depicted in Fig.~\ref{fig:energy_level}. The blue energy levels consist of the pseudo-spin-1 system as mentioned in the main article, and the solid blue arrows represent the radio frequency and microwave drives. The dashed black arrows illustrate the shelving path where we transfer population of state $\ket{\psi_1}$ ($\ket{\psi_3}$) to the very dark state $\ket{1,-1}$ ($\ket{2,-2}$) when detecting the population of state $\ket{\psi_2}$.

For each data point, the preparation, evolution and detection are repeated 500 times. We simulate the error by assuming the normal distribution of the raw data, and sample 500 times from the raw data, leading to an estimation of the standard deviation of the Berry flux. 


\begin{figure}
    \centering
    \includegraphics[width=0.6\columnwidth]{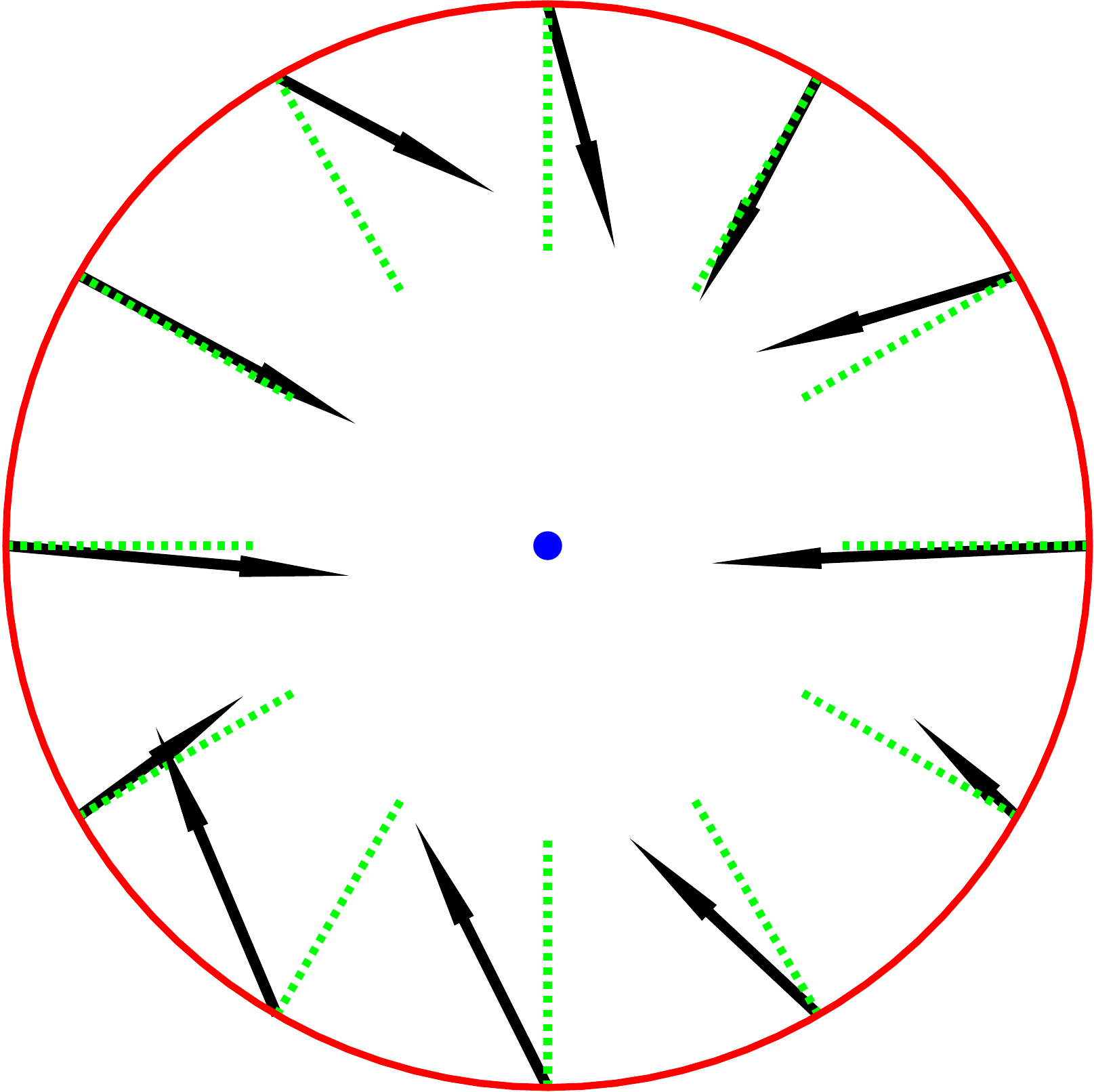}
    \caption{Spin vortex around north pole with $\alpha>1$ and $\beta=0$. The black arrows are the experimental measured spin vectors and the green dashed lines are the corresponding theoretical ones. The red circle corresponds to the chosen loop with  $\theta=0.1$ and the blue dot is the gap-closing point.}
    \label{fig:alpha_vortex}
\end{figure}

\emph{Measurement of Observables.}
In the experiment we can prepare arbitrary state from $\ket{\psi_2}$ with two pulses $\hat{R}_{12}(\theta_1,\phi_1)$ and $\hat{R}_{23}(\theta_2,\phi_2)$. Where $\hat{R}_{ij}$ is the single qubit rotation operator on corresponding subspace. Thus measuring the population of the eigenstates of $\hat{N}_{ij}$ or $\hat{F}_i$ can be realized via two analytical pulses followed by detecting the polulation of $\ket{\psi_2}$, this can be expressed as

\begin{equation}
    \langle eig|\psi_f \rangle=\bra{\psi_2}\hat{R}_{12}^\dagger\hat{R}_{23}^\dagger\ket{\psi_f}
\end{equation}
where $\ket{eig}$ is a certain eigenstate need to be detected, $\ket{\psi_f}$ is the final state of the ion and $\hat{R}_{ij}^\dagger(\theta_i,~\phi_i)=\hat{R}_{ij}(\theta_i,~\phi_i+\pi)$. 
To detect $\ket{\psi_2}$, we first shelve the other two states to the corresponding dark states as the black arrows indicate in 
Fig.~\ref{fig:energy_level}. Then we apply laser beam near $^2S_{1/2}$ to $^2P_{3/2}$ at approximately $313.349$ nm, which drives the cycling transition between $^2S_{1/2}\ket{2,2}$ and $^2P_{3/2}\ket{3,3}$, then the fluorescence emitted by the ion can be collected. 

The desired Hamiltonian $H=\mathbf{k}\cdot \hat{\mathbf{F}}+\alpha k_z \hat{N}_{zz}+\beta k_x \hat{N}_{xz} $ is implemented by detuned drives from resonances, where the detunings are temporally varied and a corresponding time-dependent rotating frame is applied following the detunings, which implements the time-dependent diagonal terms of the Hamiltonian. Therefore, for such time-dependent rotating frame, an extra phase is accumulated with respect to the resonant frame which is
\begin{equation}
    \Phi_{ij}(t) = -\int_0^t \delta_{ij}(\tau)d\tau,
\end{equation}
where $\delta_{ij}(\tau)$ is the frequency detuning of the transition from the resonant frequencies $\omega_{ij}$. For measurements where analytical pulses
are applied, we apply a compensation by shifting the initial phases of the drive with amount $(-1)^i\Phi_{ij}(t_m)$ such that the resonant drives can rotate the spin to the desired measurement basis, where $t_m$ is the measurement-performing time. We find it a crucial step to obtain correct measurements here where time-dependent detunings are utilized.

\emph{Decoherence effect and adiabatic requirement.}
To prepare ground states for the desired Hamiltonian around the exceptional point, it is ideal to evolve following an adiabatic path, which requires generally a slow evolution. However, in the experiment we need to evolve the state quick enough within the finite coherence time between each transitions. Limited by the measured coherence time between states $\ket{\psi_2}\leftrightarrow\ket{\psi_3},\ket{\psi_1}\leftrightarrow\ket{\psi_2}$ and $ \ket{\psi_1}\leftrightarrow\ket{\psi_3}$ which are 0.89(5) ms, 2.8(2) ms and 6(1) ms respectively. Experimentally we set total evolution time 1 ms that does not degrade the data in an obvious way while demonstrating the physical process. The adiabaticity is further verified with a numerical simulation, as shown in Fig.~\ref{fig:adiacon}, the bias of Berry phase is less than $0.12\pi$ 
with 1 ms ramping time.

\begin{figure}[!t]
    \centering
    \includegraphics[width=1.0\columnwidth]{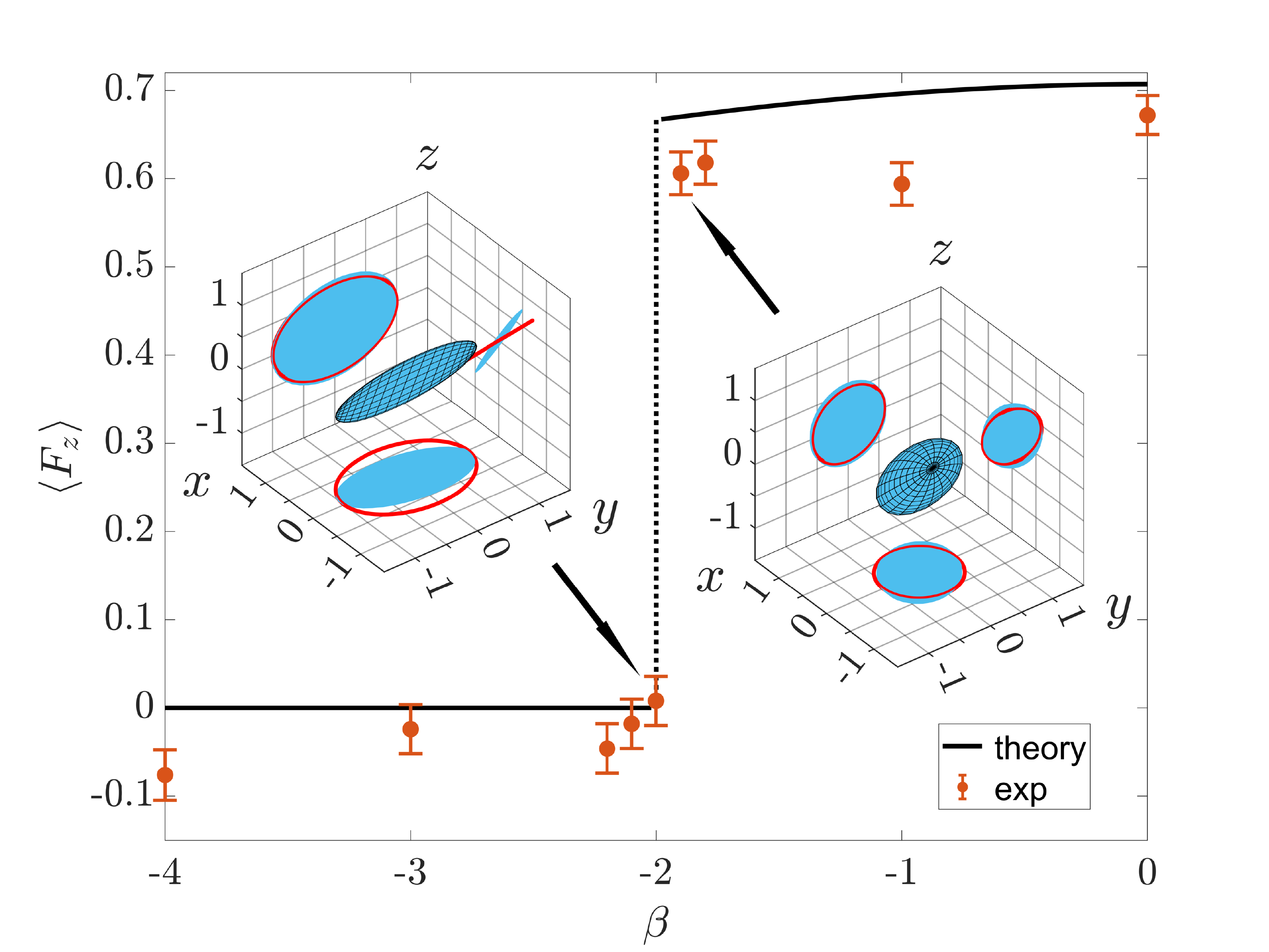}
    \caption{Phase transition characterized by the jumps of spin  vectors and tensors with $\alpha=0$. 
        The insets describe tensor ellipsoids and their projection
        on $x,~y$ and $z$ plane at $\beta=-2.001$ (left with $\mathcal{C}=0$) and $\beta=-1.9$ (right with $\mathcal{C}=2$). The red circles and lines are the projections of the theoretical tensor ellipsoids.}
    \label{fig:beta_fz}
\end{figure}

\begin{figure} 
   
	\begin{center}
    \includegraphics [width=1\columnwidth]{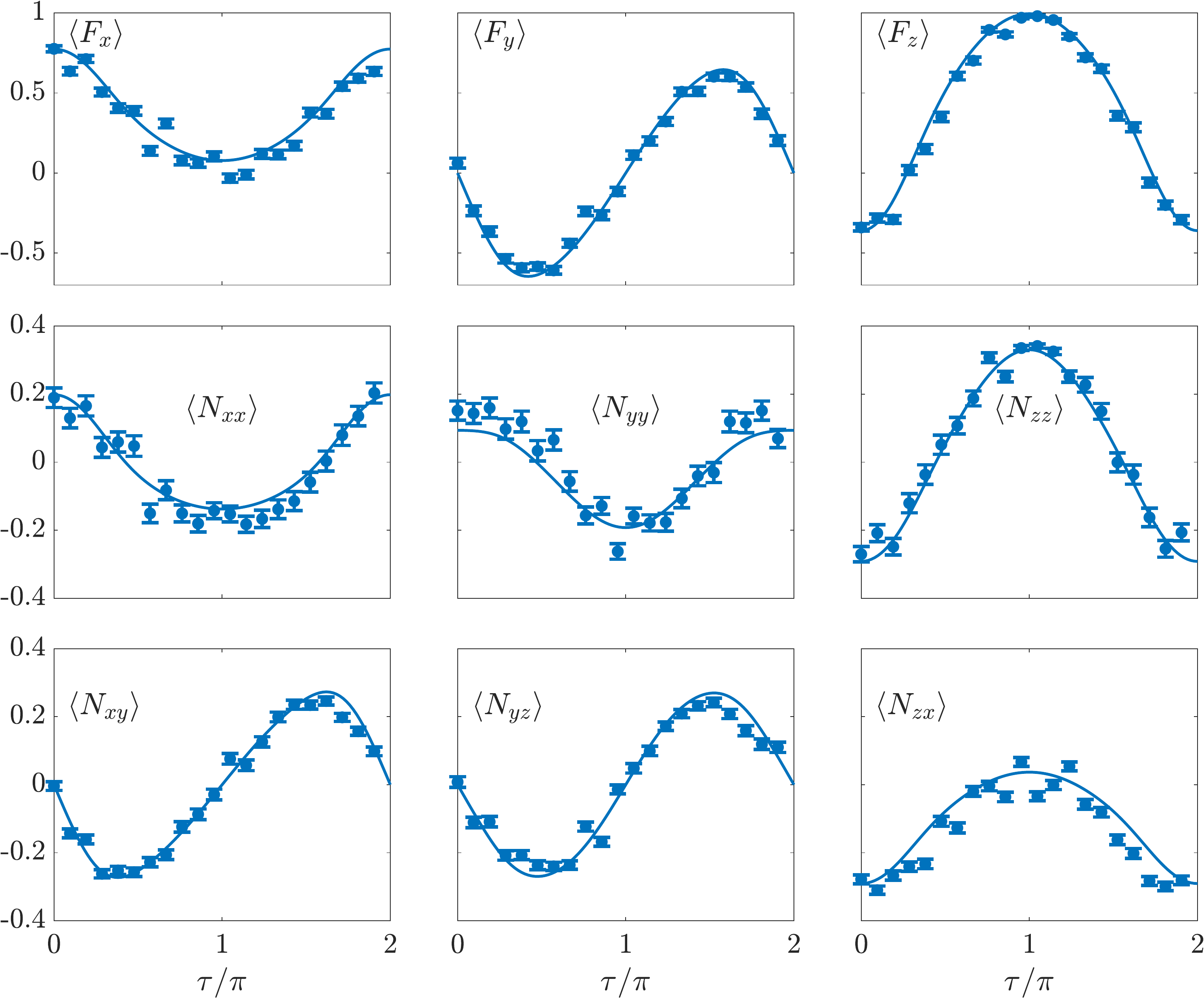}
    \caption{Sample data for the measurement of vectors and tensors with parameters $\alpha =0, \beta=-1.9, r=0.2$. The figures show the mean values of $\hat{F}_i$ and $\hat{N}_{ij}$ which are essential to reconstruct the vector arrows and tensor ellipsoids, with the solid lines representing theoretical predictions and red dots marking experimental data with error bars.}
    \label{fig:beta19_ave}
    
	\end{center}
\end{figure}

\emph{Observation of topological phase transition.}
For the transition from $\mathcal{C}=2$ to $\mathcal{C}=1$ with $\beta=0$, the gap close and reopen at the north pole $\theta=0$ of the momentum sphere at $\alpha=\alpha_c\equiv1$. After the phase transition with $\alpha>1$, spin vortex appears around the north pole as observed in our experiment.  
We choose a small latitude loop around north pole with azimuthal angle $\theta=0.1$ and measure the spin vectors along the loop with $\alpha=2$. As depicted in Fig.~\ref{fig:alpha_vortex}, we find the spin vectors roughly pointing to the north pole (gap-closing point) with $\phi_F\simeq-\phi$.

For the transition from $\mathcal{C}=2$ to $\mathcal{C}=0$, we set $\alpha=0$ and decrease $\beta$ from $0$ to $-4$. 
The band gaps close at $(\theta=3\pi/4,~\phi=\pi)$ on the momentum sphere as $\beta$ changes across $\beta_c=-2$, and jumps of both spin vector and tensors are expected at the gap-closing point. As illustrated in Fig.~\ref{fig:beta_fz}, we observe such jumps across phase transition by measure the observables $\langle \hat{F}_i \rangle$, $\langle \hat{N}_{ij}\rangle$ for $\{i,j\}=\{x,y,z\}$ along the adiabatic loop. A typical measurement result is shown in Fig. \ref{fig:beta19_ave}.

\end{document}